\documentclass[peerreview,11pt]{IEEEtran}
\usepackage[english]{babel}
\usepackage{float}
\usepackage{graphicx}
\usepackage{amsmath}
\usepackage{amsfonts}
\usepackage{amssymb}
\usepackage[utf8]{inputenc}
\usepackage{lmodern}
\usepackage{verbatim}
\usepackage{cite} 
\usepackage{amsthm}
\usepackage{caption}
\usepackage{subcaption}
\usepackage[colorlinks=true, allcolors=blue]{hyperref}

\newtheorem{theorem}{Theorem}[section]

\newtheorem{proposition}[theorem]{Proposition}

\theoremstyle{definition}
\newtheorem{definition}[theorem]{Definition}
\newtheorem{remark}[theorem]{Remark}
\newtheorem{example}[theorem]{Example}

\newcommand{\R}{\ensuremath{\mathbb{R}}}

\newcommand{\F}{\ensuremath{\mathbb{F}}}

\usepackage[square, authoryear]{natbib}
\numberwithin{equation}{section}
\usepackage{authblk}

\begin{document}
\title{
Completeness of Riemannian metrics: an application to the control of constrained mechanical systems}

\author{Jos\'e \'Angel Acosta, Anthony Bloch and David\ Mart\'{\i}n de Diego
\thanks{Jos\'e \'Angel Acosta is with Department of Ingenier{\'\i}a de Sistemas y Autom\'atica, Universidad de Sevilla, Sevilla, 41092, Spain (e-mail: jaar@us.es).}
\thanks{Anthony Bloch is with Department of Mathematics, University of Michigan 530 Church Street, Ann Arbor, MI, USA (e-mail: abloch@umich.edu).}
\thanks{David\ Mart\'{\i}n de Diego is with Instituto de Ciencias Matem\'aticas (CSIC-UAM-UC3M-UCM), C/Nicol\'as Cabrera 13-15, 28049 Madrid, Spain (e-mail: david.martin@icmat.es).}}

%
%

%
\date{\today}

\markboth{J.\'{A} Acosta, A. Bloch and D. Mart\'{\i}n de Diego. Completeness of Riemannian metrics}%
{}

\maketitle

\begin{abstract}
We introduce a mathematical technique based on modifying a  given Riemannian 
metric and we investigate its applicability to controlling and stabilizing constrained mechanical systems. In essence our result is based on the construction of a complete Riemannian metric in the modified space where the constraint is included. 
%
%
In particular this can be applied to the controlled Lagrangians technique \cite{controlled-1,controlled-2} modifying its metric to additionally cover mechanical systems with configuration constraints via control.
The technique used consists of approximating  incomplete Riemannian metrics by complete ones, modifying the evolution
near a boundary and finding a controller satisfying a given design criterion.
 \end{abstract}	



\section{Introduction}
	The study of maximal intervals of existence  of solutions of ordinary differential equations is a classical problem in Mathematics (see, for instance, \cite{Hartman}).
 This study was  extended to the case of second order differential equations on Riemannian manifolds 
(see \cite{We-Mar-70, CaRoSa,Minguzzi})  and, in particular, to the geodesic sprays of Riemannian metrics. 
We show that these results combined with classical results about approximation of incomplete Riemannian metrics by complete ones \cite{Fegan-Millman,nomizu-ozeki,Gordon,Gordon2},
provide us with a novel and useful technique with applications to control problems with state space constraints such as collision avoidance. Typical applications with these constraints range from the analysis of robotic systems with joint limits, avoiding self-collisions and/or collisions with other external obstacles, to controlling the course of unmanned vehicles in confined environments. 

In a recent work \cite{Klein23} the authors propose  a modification of the initial Riemannian metric allowing for the generation of robot motions as geodesics that efficiently avoid given regions. Our work provides a complementary mathematical formalism and solution that develops this theory by connecting with completeness results of Riemannian metrics. In our work we only consider modifications that completely comply with constraints/restrictions on motion, but other modifications of the Riemannian metric that are less strict and more easily realisable in practical applications could be considered, so-called soft constraints.

In a different context of the control of constrained nonlinear systems it is worth mentioning some promising lines of research based on either barrier certificates and functions or diffeomorphisms. The barrier certificates are used in  verifying that trajectories do not enter in unsafe regions see e.g. \cite{Prajna2007,wieland2007} and rely on optimization techniques. The second approach is based on the construction of control Lyapunov functions which include barriers see e.g. \cite{Tee2009,romdlony2016}. Finally, closely related to differential geometry, there are approaches based on the construction of diffeomorphisms that map all the trajectories of the constrained dynamics into unconstrained dynamics as in e.g. \cite{Acosta2018}. All these lines of research share the construction of some kind of barrier, however none takes the advantage of the Riemannian metrics of mechanical systems.

In summary, we provide two main contributions in Theorems \ref{teore} and \ref{teo-2}. On the one hand, 
in Theorem~\ref{teore} we give a characterization of a control law for underactuated mechanical systems which by construction avoids an obstacle defined by a function in the configuration space. On the other hand, in Theorem~\ref{teo-2} we complement the classical result of controlled Lagrangians \citep{controlled-1,controlled-2} for stabilization of underactuated mechanical systems with configuration unilateral constraints (see e.g. \cite{Brogliato}). Numerous practical applications, such as robotic control, air vehicle control, marine surface vessel control, etc. typically require these types of unilateral constraints in their modeling and control. We provide some applications of the previous result to different examples including the classical pendulum on a cart.


The paper is organized as follows: In Section \ref{riemannian-section}, we review some results on complete and incomplete Riemannian metrics. In Section \ref{sode-section} we consider the extension to the case of mechanical systems, adding to a complete  Riemannian metric appropriate potential and external forces. In Section \ref{section:control} we prove one of the main results of the paper (Theorem \ref{teore}), where we show how to derive trajectories of a given control system that avoid a region determined by unilateral constraints via a modification of the Riemannian metric so that it is complete.
We conclude this section by giving some simple examples that demonstrate the usefulness of our technique for avoiding different types of obstacles. 
Finally, in Section \ref{sec:constrained-control} we derive a  modification of the stabilization techniques given in \cite{controlled-1,controlled-2} to derive a  control strategy that additionally allows one  to confine the motion of the
system to a prescribed region determined by unilateral constraints. We demonstrate the applicability of these results  for the stabilization of a constrained pendulum on a cart.

	\section{Complete and incomplete Riemannian metrics}\label{riemannian-section}
	
 Consider a finite dimensional smooth manifold $Q$, $\dim Q=n$, 
 denote by $TQ$ the  tangent bundle of the manifold (phase space of positions and velocities) and $\tau_Q: TQ\rightarrow Q$ the canonical projection defined by  $\tau_Q(v_q)=q$ for all $v_q\in T_q Q$ \cite{AbMa}. 
 Given a wvector field $X\in {\mathfrak X}(Q)$, we say that $X$ is {\bf complete} if every one of its flow curves exist for all time. That is, if for any initial condition $q_0\in Q$ there exists an integral  curve  of $X$, $\sigma: (-\infty, \infty)\rightarrow Q$ such that $\sigma(0)=q_0$. 
 We are particularly interested in Riemannian geodesics (see \cite{doCarmo}). Similarly, we say that a Riemannian manifold  $(Q,g)$ is (geodesically) complete if for all $p\in Q$ any geodesic $\gamma(t)$ starting from $p$ is defined for all values of the parameter $t$. 
	
	\begin{example}
	Common examples of completeness: The Euclidean space with the usual metric is complete, but the open unit disk with the standard metric is not complete since any geodesic ends at a finite time. 
	
	 The punctured plane ${\mathbb R}^2\setminus \{(0,0)\}$ is not complete with the usual metric.
	However, with the Riemannian metric $ds^2=\frac{1}{r^2}(dx^2+dy^2)$, the punctured plane is identified with the infinite (flat) cylinder, which is complete. 
	\end{example}

	The following alternative characterizations of completeness for Riemannian metrics will be useful (see for instance \cite{Milnor}): 
	\begin{proposition}
	Let $(Q, g)$ be a connected Riemannian manifold. The following statements are equivalent:
	\begin{enumerate}
		\item  The manifold $(Q, g)$ is geodesically complete; that is, for every $p\in Q$, every geodesic through $p$ can be extended to a geodesic defined on all of ${\mathbb R}$.
	\item For every point, $p\in Q$, the map $exp_p$ is defined on the entire tangent space, $T_pQ$.
	\item Any closed and bounded subset of the metric space $(Q, d)$ is compact ($d$ denotes the Riemannian distance). 
	\item  The metric space $(Q, d)$ is complete (that is, every Cauchy sequence converges).
	\end{enumerate}
	\end{proposition}

	The following result proved in \cite{Fegan-Millman} is instrumental for our technique of control constrained mechanical systems since we can approach any incomplete metric using complete metrics.
	\begin{theorem}\label{teo1}
	Let $g_0, g_1$ Riemannian metrics on $Q$. 
If $g_0$ is a complete Riemannian metric then for all $s>0$ and $t\geq 0$
	\[
	sg_0+t g_1
	\]
	is also a complete metric. 
	\end{theorem}
In \cite{Fegan-Millman}, the authors comment as an interpretation of this theorem, that the incomplete metrics must lie on the edge of the set of all Riemannian metrics. 

In Theorem \ref{teo-principal} later in the manuscript we use the following result.
\begin{theorem}\label{gordon-theorem}[\cite{Gordon,Gordon2}]
If $g$ is a Riemannian metric on $Q$ (not necessarily complete) then for any proper  function on $Q$, $f:Q\rightarrow {\mathbb R}$, the metric $\tilde{g}=g+df\otimes df$ is complete. 
\end{theorem}
Let clarify what we mean by a proper function.  A proper function is proper if inverse images of compact subsets are compact. In particular, if $Q$ is an open domain with compact closure $\delta Q$, then a function $f: Q\rightarrow {\mathbb R}$ is proper if $|f(q)|\rightarrow \infty$ as $q\rightarrow \delta Q$. 

As we know the Riemannian geodesics are characterized as the solutions of the Euler-Lagrange equations for the Lagrangian $K_g: TQ\rightarrow {\mathbb R}$ given by $K_g(v_q)=\frac{1}{2}g_q(v_q, v_q)$ or in coordinates
\[
K_g(q^i, \dot{q}^i)=\frac{1}{2}g_{ij}\dot{q}^i\dot{q}^j,
\]
where $g=g_{ij}d q^i\otimes dq^j$. 
The metric is complete in $Q$ adding  a proper function as in Theorem \ref{gordon-theorem} and modifying the Lagrangian system using the new Lagrangian system $K_{\tilde{g}}$ locally expressed as 
\[
K_{\tilde{g}}(q^i, \dot{q}^i)=\frac{1}{2}(g_{ij}+f_i f_j)\dot{q}^i\dot{q}^j,
\]
where $f_i=\partial f/\partial q^i$, $1\leq i\leq \dim Q$. 

Another interesting result proved by \cite{nomizu-ozeki} is stated in the following theorem.
\begin{theorem}
For any Riemannian metric $g$ on $Q$, there exists a
complete Riemannian metric which is conformal to $g$.
\end{theorem}

\section{Completeness for mechanical systems}\label{sode-section}


We describe a mechanical system using the following three ingredients from \cite{Bullo-Lewis}: 
\begin{itemize}
\item $g$ is a Riemannian metric on a manifold $Q$.
Let denote by 
$\flat_g: TQ\rightarrow T^*Q$ the vector bundle isomorphism defined by 
$\langle \flat_g (v_q), w_q)=g_q(v_q,w_q)$, for all $v_q, w_q\in T_qQ$ and 
its inverse by $\sharp_g=(\flat_g)^{-1}: T^*Q\rightarrow TQ$.
The Riemannian metric induces  the kinetic energy $K_g$ on $TQ$ 
by $K_g(v_q)=\frac{1}{2} g_q(v_q, v_q)$ (see Section \ref{riemannian-section}).

\item The potential energy function $V\in C^{\infty}(Q)$.
\item Non-conservative forces defined as a map $F: TQ\rightarrow T^*Q$ such that $\pi_Q \circ F = \tau_Q$  
where $\pi_Q: T^*Q\rightarrow Q$ is the canonical projection given by $\pi_Q(\mu_q)=q$ for all $\mu_q\in T^*_q Q$. 
In canonical bundle coordinates $(q^i, \dot{q}^i)$ and $(q^i, p_i)$ on $TQ$ and $T^*Q$, respectively,  we have $F(q^i, \dot{q}^i)=(q^i, p_i=F_i(q, \dot{q}))$. 
\end{itemize}
Therefore,  a mechanical system is defined  by $(L=K_g-V, F)$. 
Observe that in local coordinates $(q^i, \dot{q}^i)$ in $TQ$ 
\[
L=\frac{1}{2}g_{ij} \dot{q}^i\dot{q}^j-V(q) ,
\]
where $g=g_{ij}dq^i\otimes dq^j$. 
%
%
The solutions of the mechanical system $(L, F)$ are the  curves $c: I\subset \R\rightarrow Q$ such that  
\begin{equation}\label{forced}
\nabla^{g}_{\dot{c}(t)}\dot{c}(t) +\hbox{grad}_g V(c(t))=Y_F(c(t), \dot{c}(t))\; .
\end{equation}
where $\nabla^{g}$ is the Levi-Civita connection, $g(q)(\hbox{grad}_g V(q), \cdot)=dV (q)$ and $g(q)(Y_F(v_q), \cdot)=F(v_q)$ for all $q\in Q$ (see, for instance, \cite{Bullo-Lewis}), which in coordinates reads
\[
\ddot{q}^k+\Gamma^k_{ij}\dot{q}^i\dot{q}^j+g^{ki}\frac{\partial V}{\partial q^i}=g^{ki}F_i, \quad 1\leq k\leq n,
\]
where  $\Gamma^{k}_{ij}$ are the Christoffel symbols of the Levi-Civita connection $\nabla^g$ (that is
$\nabla^g_{\partial_i}\partial_j=\Gamma_{ij}^k\partial_k$ where $\partial_i$ is a shorthand notation for $\frac{\partial}{\partial q^i}$). 
Observe that $\hbox{grad}_g V(q)=g^{ki}\frac{\partial V}{\partial q^k}\frac{\partial}{\partial q^i}$ and $Y_F=g^{ki}F_k\frac{\partial}{\partial q^i}$ is a   vector field along the projection $\tau_Q: TQ\rightarrow Q$.  

The equations (\ref{forced}) are equivalent in local coordinates to the forced Euler-Lagrange equations given by
\[
\frac{d}{dt}\left( \frac{\partial L}{\partial \dot{q}^i}\right)-\frac{\partial L}{\partial {q}^i}=F_i, \quad 1\leq i\leq n=\dim Q\; .
\]

Additionally, the force $F$ is {\bf weakly dissipative}  if $\langle dK_g, Y^v_F\rangle\leq 0$ and {\bf dissipative} if 
$\langle dK_g, Y^v_F\rangle (v_q)<0$ for all  $v_q\in T_qQ$ such that $v_q\not=0_q$, where $Y_F^v\in {\mathfrak X}(TQ)$ is the vertical lift of $Y_F$. Locally $Y^v_F=g^{ki}F_k\frac{\partial}{\partial \dot{q}^i}$ and
\[
\langle dK_g, Y^v_F\rangle=F_i(q, \dot{q})\dot{q}^i\leq 0.
\]

In the following definition, we characterize when a mechanical system is complete. 

\begin{definition} (see \cite{AbMa,We-Mar-70})
A function $V_0: [0, +\infty)\rightarrow \R$ is called {\bf positively complete} if it is $C^1$, non-increasing and satisfies
\[
\int_0^{+\infty} \frac{ds}{\sqrt{e-V_0(s)}}\, ds =+\infty,
\]
where $e$ is a constant with $e>V_0(s)$ for all $s\in[0,+\infty)$.
\end{definition}
As examples of functions $V_0$ verifying such property we have $V_0(x)=-x^{\alpha}$ with $0\leq \alpha\leq 2$ or  $V_0(x)=-x (\log(1+x))^{\beta}$ with $\beta>0$.

However, in our approach it will only be necessary to characterize when the solutions are forward complete (or positively  complete), that is,  if $c(0)$ and $\dot{c}(0)$ are initial conditions then the solutions are defined on $[0, +\infty)$.
From \cite{AbMa} (Proposition 3.7.18) we have the following result.

\begin{theorem}\label{teo-principal}
    Let $(M, g)$ be a complete Riemannian manifold, $F: TQ\rightarrow T^*Q$ a dissipative  external force and $V: Q\rightarrow {\mathbb R}$ a potential function. Assume 
    there is a positively complete function $V_0$ such that 
    \begin{equation}\label{posi}
    V_0(d_g(q, q_0))\leq V(q)
    \end{equation}
 for  some $q_0$ and where $d_g$ is the distance between $q$ and $q_0$.  Then the solutions of the system (\ref{forced})
    are forward  complete.
\end{theorem}

For example if $-V$ grows at most quadratically, that is there are a $q_0\in Q$ and constants $a,b>0$ such that
\[
-V(q)\leq a+b(d_g(q_0, q))^2
\]
(this property is independent of the chosen $q_0$) and if $F$ is dissipative then the flow is backward complete.  
%
A specific example is given by the force $F(q^i, \dot{q}^i)=-\sum_{i=1}^n k_i\dot{q}^i$ with $k_i\geq 0$ which is dissipative in $({\mathbb R}^n, g_{eucl})$.


Let consider the system with Lagrangian $L(x, \dot{x})=\frac{1}{2}\dot{x}^2$,  $V(x)=-\frac{1}{2} x^{2(1+\epsilon)}$ with $\epsilon>0$ and $F\equiv 0$ (see \cite{CaRoSa2013}), then the equations of motion are
$
\ddot{x}=(1+\epsilon)x^{1+2\epsilon}
$.
For initial conditions $x(0)=1$ and $\dot{x}(0)=0$ we derive from the preservation of energy that
\[
\frac{1}{2}\dot{x}^2+\frac{1}{2} x^{2(1+\epsilon)}=\frac{1}{2},
\]
which we can solve for $t$ and deduce
\[
t(x)=\int_1^{x}\frac{ds}{\sqrt{s^{2(1+\epsilon)}-1}}<\infty,
\]
and therefore the system is not complete. 

\begin{remark}
In \cite{CaRoSa2013} the authors study the completeness of solutions of more general second order differential equations, in particular, they include explicitly time-dependent systems. This results will be useful for future extensions of our framework (see Section \ref{section:conclusions}).
\end{remark}

\section{Control of constrained mechanical systems}\label{section:control}

Let $Q$ be the configuration space equipped with a $C^3$-Riemannian metric $g$. Denote by $M=\{ q\in Q\; |\; \phi(q)<0\}$ the open subset defined by a smooth function $\phi: Q\rightarrow \R$.
Consider the mechanical control system 
\begin{equation}\label{control}
\nabla^g_{\dot{c}(t)}\dot{c}(t)+\hbox{grad}_g V(c(t))=u^a(t) Y_a(c(t)),
\end{equation}
where $Y_a$, $1\leq a\leq m$ with $m\leq \dim Q$ are the control vector fields and $u^a$ plays the role  of the control inputs \citep{Bullo-Lewis}.
In coordinates it reads
\[
\ddot{q}^k+\Gamma^k_{ij}\dot{q}^i\dot{q}^j+g^{ki}\frac{\partial V}{\partial q^i}=u^a Y_a^k, \quad 1\leq k\leq n,
\]
where $Y_a=Y_a^k\frac{\partial}{\partial q^k}$.
Denote by 
$
{\mathcal D}_c=
\hbox{span }\{ Y_a, 1\leq a\leq m\}
$,
the control distribution. Alternatively, we initially can have control forces given by a set 
$
{\mathcal F}_c=
\hbox{span }\{ \mu_a, 1\leq a\leq m\}\subseteq \Lambda^1 Q.
$
In this case, we have the relation $\sharp_g({\mathcal F}_c)={\mathcal D}_c$ where now $Y_a=\sharp_g(\mu_a), 1\leq a\leq m$. Locally if $\mu_a=(\mu_a)_i(q)\, dq^i$ then 
$Y_a=g^{ij}(\mu_a)_i\frac{\partial}{\partial q^j}$.

In the sequel we will  assume that $g$ is a complete Riemannian metric on $Q$ and the potential function $V$ satisfies (\ref{posi}). Consequently the second order system defined by the regular Lagrangian system $L(v_q)=\frac{1}{2}g(v_q,v_q)-V(q)$ is complete on $Q$. In the following theorem we state one of our contributions.

\begin{theorem}\label{teore}
Given the mechanical control system (\ref{control}) and the open subset $M=\{ q\in Q\; |\; \phi(x)<0\}\subseteq Q$, if we find a proper map $f: M\rightarrow \R$ such that $\hbox{grad}_g f\in {\mathcal D}_c$ (or $df\in {\mathcal F}_c$) then 
the solutions of (\ref{control}) with control input 
\begin{equation}\label{equation:control}
u_*^a={\mathcal C}^{ab}(1+|\hbox{grad}_g f|^2)^{-1}
\left(
f^j\frac{\partial V}{\partial q^j}-f_{jk}\dot{q}^j\dot{q}^k\right)\langle df, Y_b\rangle
\end{equation}
and initial condition $c(0)\in M$ are complete in $M$, where the following notation is in order
\begin{eqnarray*}
f_i&=&\frac{\partial f}{\partial q^i},\\
f^i&=& g^{ik}\frac{\partial f}{\partial q^k},\\
|\hbox{grad}_g f|^2&=&\langle df, \hbox{grad}_g f\rangle
=g^{ij}\frac{\partial f}{\partial q^i}
\frac{\partial f}{\partial q^j}=f^i f_i,\\
f_{jk}&=&\frac{\partial^2 f}{\partial q^j\partial q^k}-\Gamma_{jk}^i\frac{\partial f}{\partial q^i},\\
{\mathcal C}_{ab}&=&g(Y_a, Y_b), \qquad ({\mathcal C}^{ab})=({\mathcal C}_{ab})^{-1}\; .
\end{eqnarray*}
\end{theorem}
\begin{proof}
Since $f: M\rightarrow \R$  is proper then, from  Theorem \ref{gordon-theorem}, we have that
$\tilde{g}=g+df\otimes df$ is complete in $M$. Define the mechanical Lagrangian $L_f: TM\rightarrow \R$ by 
\[
L_f( v_q)=\frac{1}{2}g(v_q, v_q)+\frac{1}{2}\langle df(q), v_q\rangle^2-V(q)
\]
where $q\in M$ and $v_q\in T_q M$. Then, from Theorem \ref{teo-principal}, the system $(\tilde{g}, V_M)$
is complete where $V_M: M\rightarrow \R$ is the restriction of $V$ to $M$. 

It is straightforward to check that (see \cite{Gordon,Gordon2,Gordon2-correction})
\begin{eqnarray*}
\tilde{g}_{ij}&=& g_{ij}+f_i f_j,\\
\tilde{g}^{ij}&=&g^{ij}-(1+|\hbox{grad}_g f|^2)^{-1} f^i f^k\\
\tilde{\Gamma}^k_{ij}&=&{\Gamma}^k_{ij}+(1+|\hbox{grad}_g f|^2)^{-1}f_{ij}f^k
\end{eqnarray*}
Therefore the equations for the complete system given by $L_f: TM\rightarrow \R$ are
$$
\ddot{q}^k  + \tilde{\Gamma}^k_{ij}\dot{q}^i\dot{q}^j+\tilde{g}^{ki}\frac{\partial V}{\partial q^i}=0
$$ or in terms of the original mechanical system read
$$
\ddot{q}^k + {\Gamma}^k_{ij}\dot{q}^i\dot{q}^j+{g}^{ki}\frac{\partial V}{\partial q^i}=  
(1+|\hbox{grad}_g f|^2)^{-1}\left(
  f^j\frac{\partial V}{\partial q^j}-f_{ij}\dot{q}^i\dot{q}^j\right)f^k \; .
$$
    Now since   $\hbox{grad}_g f\in {\mathcal D}_c$ 
    then $\hbox{grad}_g f=\lambda^a Y_a$, $1\leq a\leq m$. Therefore
    $$
    \langle df, Y_b\rangle= g(\hbox{grad}_g f, Y_b)=\lambda^a g(Y_a, Y_b)$$ and we conclude that
    $$\hbox{grad}_g f={\mathcal C}^{ab} g(\hbox{grad}_g f, Y_b) Y_a={\mathcal C}^{ab} \langle df, Y_b\rangle Y_a .$$
\end{proof}

Alternatively, for the last part of the proof above, we can rewrite the equations of motion as
\begin{equation}\label{eq:alter-1}
\ddot{q}^k + {\Gamma}^k_{ij}\dot{q}^i\dot{q}^j+{g}^{ki}\frac{\partial V}{\partial q^i}=u^a_*Y_a^k  
\end{equation}
with controls given by (\ref{equation:control}) as follows
\begin{eqnarray}
(\mu_a)_k\left(\ddot{q}^k + {\Gamma}^k_{ij}\dot{q}^i\dot{q}^j+{g}^{ki}\frac{\partial V}{\partial q^i}\right)&=&u^*_a={\mathcal C}_{ab}u^b_*,
\label{eq:alterna-2a}\\ 
  (\mu_\alpha)_k\left(\ddot{q}^k + {\Gamma}^k_{ij}\dot{q}^i\dot{q}^j+{g}^{ki}\frac{\partial V}{\partial q^i}\right)&=&0,  \label{eq:alterna-2b}
\end{eqnarray}
where
$$
u^*_a={\mathcal C}_{ab}u^b=(1+|\hbox{grad}_g f|^2)^{-1}
\left(
f^j\frac{\partial V}{\partial q^j}-f_{jk}\dot{q}^j\dot{q}^k\right)\langle df, Y_a\rangle
$$
using a basis of 1-forms $\{\mu_a, \mu_{\alpha}\}$ where $\mu_a=\flat_g(Y_a)$, $1\leq a\leq m$, and $\mu_{\alpha}$, $m+1\leq \alpha\leq n$, is a basis of the annihilator ${\mathcal D}_c^0$ of the control distribution. 
Observe that $\mu_a (Y_b)= {\mathcal C}_{ab}$ and $\mu_\alpha (Y_b)=0$.
\begin{remark}
Other possibilities in Theorem \ref{teore} like the presence of external forces or a time-dependent proper function,  $f_t: M_t\rightarrow {\mathbb R}$, are trivial but interesting in applications specially in avoiding moving obstacles.  
\end{remark}
\begin{remark}\label{remark-1}
For computational purposes is convenient, in some case, to use a covariant version of Theorem~\ref{teore}. 
To this end, we start with a Lagrangian function $L: TQ\rightarrow {\mathbb R}$ and we modify the dynamics adding an additional Lagrangian $\widetilde{L}: TQ\rightarrow {\mathbb R}$ in such a way that the sum
$\mathbf{L}=L+\widetilde{L}$ is a regular Lagrangian. 
Let us denote by $(W_{ij})$, $(\widetilde{W}_{ij})$ and $(\mathbf{W}_{ij})$ the hessian matrices of $L$, $\widetilde{L}$ and $\mathbf{L}$, respectively. 
From the Euler-Lagrange equations for $\mathbf{L}$ we get
\[
\frac{d}{dt}\left( \frac{\partial \mathbf{L}}{\partial \dot{q}^i}\right)-
\frac{\partial \mathbf{L}}{\partial q^i}=0
\]
and deduce that 
\[
\ddot{q}^j=-\mathbf{W}^{ij}\left(
\frac{\partial^2 \mathbf{L}}{\partial \dot{q}^i\partial \dot{q}^k}\dot{q}^k-\frac{\partial \mathbf {L}}{\partial q^i}\right).
\]
Therefore, using that $\mathbf{L}=L+\widetilde{L}$ we have that these equations are equivalent to 
\begin{eqnarray*}
\frac{d}{dt}\left( \frac{\partial {L}}{\partial \dot{q}^i}\right)-
\frac{\partial {L}}{\partial q^i}&=&
-\widetilde{W}_{ij}\ddot{q}^j-\frac{\partial^2 \widetilde{L}}{\partial \dot{q}^i\partial \dot{q}^k}\dot{q}^k+\frac{\partial \widetilde{L}}{\partial q^i}\\
&=&\widetilde{W}_{ij} \mathbf{W}^{jl}\left(
\frac{\partial^2 \mathbf{L}}{\partial \dot{q}^l\partial \dot{q}^k}\dot{q}^k-\frac{\partial \mathbf {L}}{\partial q^l}\right)-\frac{\partial^2 \widetilde{L}}{\partial \dot{q}^i\partial \dot{q}^k}\dot{q}^k+\frac{\partial \widetilde{L}}{\partial q^i}\; . 
\end{eqnarray*}
The calculation in Theorem \ref{teore} can be simplified using this remark with  the Lagrangian 
$
\mathbf{L}=L+\widetilde{L}$ where 
$\widetilde{L}=df\otimes df$.

\end{remark}
\subsection{Some  illustrative examples of application of completeness of trajectories  for simple control problems}

We illustrate our theory   with some  simple applications of Theorem \ref{teore}.

\begin{example}{\bf Landing-type trajectory}.
 Consider the following simple control  equations in ${\mathbb R}^2$
\[
\ddot{x}=0, \qquad \ddot{y}=u-g,
\]
where the control vector field is $\frac{\partial}{\partial y}$.
Define the feasible region  $M=\{y\in \R\; |\; y>0\}$
and consider the proper map
$f(y)=\ln y$. Then  a complete metric on $\R\times M\subset \R^2$ is
\[
dx\otimes dx +dy\otimes dy+df\otimes df=
dx\otimes dx +\left(1+\frac{1}{y^2}\right)dy\otimes dy.
\]


The dynamics to avoid the ``obstacle" $y=0$ is determined by the solutions of Lagrangian system
\[
L_f (x, y, \dot{x}, \dot{y})=\frac{1}{2} \left[\dot{x}^2+\left(1+\frac{1}{y^2}\right)\dot{y}^2\right]-g y.
\]
Theorem \ref{teore} (or Remark \ref{remark-1}) allows us to explicitly deduce the control input as 
\[
u(t)=\frac{g y+\dot{y}^2}{ y^3+y}
\]
\end{example}
\begin{figure}[h]
	\centering
\includegraphics[width=0.8\textwidth]{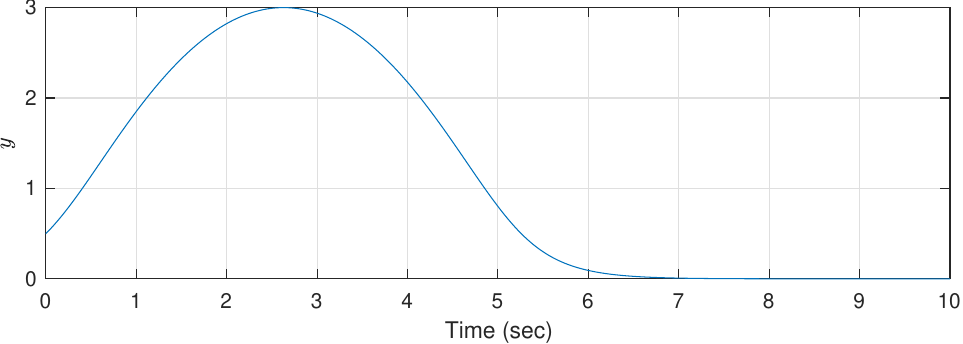}
\caption{Trajectory with initial conditions $x(0)=0.5, y(0)=0.5, \dot x(0)=0.5, \dot y(0)=1$; $\kappa=10^4$.}
\label{figure-1}
\end{figure}
A typical trajectory for the $y$ coordinate is shown in the simulation of Fig. \ref{figure-1}.

\begin{example}{\bf  Particle on a band of ${\mathbb R}^2$.}\label{example-bou}
As it is well known in non-Euclidean geometry, the Poincar\'e half-plane model is described by  the space
$
{{\mathbb H}}=\{ (x , y)\in {\mathbb R}^2\;  |\;  y>0 \}
$
 together with a complete metric, the {\bf Poincar\'e metric}, given by
 \[
 g_{Poi}=\frac{1}{y^2}(dx\otimes dx +dy\otimes dy)\; .
 \]
The geodesics are represented by circular arcs perpendicular to the $x$-axis and straight vertical rays perpendicular to the $x$-axis. 
Of course the Euclidean metric $g_{eucl}=dx\otimes dx +dy\otimes dy$ is not complete on ${\mathbb H}$. 
We will use the Poincar\'e metric for collision avoidance with the boundary of ${\mathbb H}$. 
Observe that in the previous example the velocity  in the normal component is decreasing. Now we want to obtain a type of ``bouncing motion" using the Poincar\'e metric. 
For simplicity, we assume a fully actuated control system
\begin{equation}\label{controlled}
\ddot{x}=u_1\; ,\qquad \ddot{y}=u_2,
\end{equation}
%
%
The control to avoid the constraint is constructed through the complete Lagrangian system
\[
L_{P}(x, y, \dot{x}, \dot{y})=\frac{\dot{x}^2+\dot{y}^2}{\kappa y(t)^2}-\frac{1}{2}\left(\dot{x}^2+\dot{y}^2\right)
\]
with the constant value $\kappa$ to be specified by the controller. 
Observe that this modified metric is  positive definite for  $y<\sqrt{2/\kappa}$.
Thus, the Euler-Lagrange equations for $L_P$ are
\[
y \left(\kappa y^2-2\right) \ddot{x}+4 \dot{x} \dot{y}=0, \qquad  y \left(\kappa y^2-2\right) \ddot{y}-2 \dot{x}^2+2 \dot{y}^2=0,
\]
that exactly corresponds  to  using the controls
\[
u_1=-\frac{4 \dot{x} \dot{y}}{y\left(\kappa y^2-2\right)}, \qquad u_2=\frac{2 \left(\dot{x}^2-\dot{y}^2\right)}{y \left(\kappa y^2-2\right)}.
\]
in equations (\ref{controlled}) to avoid the boundary constraint. A simulation is shown in Fig. \ref{bouncing}.
\begin{figure}[h]
	\centering
	\includegraphics[width=0.9\textwidth]{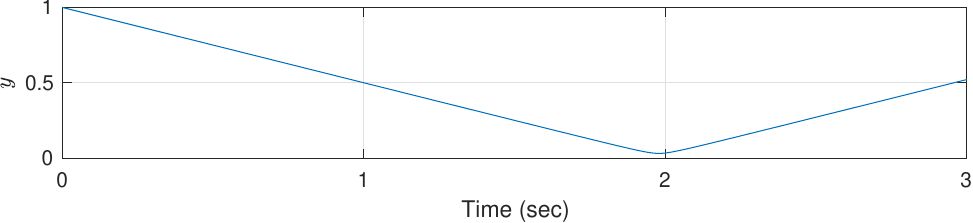}
\caption{Trajectory with initial conditions $x(0)=0, y(0)=1, \dot x(0)=1, \dot y(0)=-0.5$; $\kappa=10^4$.}
\label{bouncing}
\end{figure}


Finally, we will use modifications of the  Poincar\'e  metric to maintain the particle in the strip
$
M= \{ (x , y)\in {\mathbb R}^2\;  |\;  0<y<1 \}.
$
For that we use the following Lagrangian
\begin{eqnarray*}
L_{P'}&=&\frac{\dot x^2+ \dot y^2}{ \kappa y^2}+\frac{\dot x^2+\dot y^2}{ \kappa (y-1)^2}-\frac{1}{2} \left(\dot x^2+\dot y^2\right)
\end{eqnarray*}
with constant $\kappa>0$ that correspond to the controls for equations (\ref{controlled}) given by
\begin{eqnarray*}
u_1&=& -\frac{4 \dot{x} \dot{y} \left(\frac{1}{\kappa y^3}+\frac{1}{\kappa (y-1)^3}\right)}{-\frac{2}{\kappa y^2}-\frac{2}{\kappa (y-1)^2}+1},\\
u_2&=&\frac{2 \left(2 \kappa y^3-3 \kappa y^2+3 \kappa y- \kappa \right) \left(\dot{x}^2-\dot{y}^2\right)}{(y-1) y \left(\kappa^2 y^4-2 \kappa^2 y^3+(\kappa (\kappa-2)-2 \kappa) y^2+4 \kappa y-2 \kappa \right)}.
\end{eqnarray*}
The trajectories have to move in the strip as shown in the simulation of Fig. \ref{Figure22}.
\begin{figure}[h]
	\centering
	\includegraphics[width=0.9\textwidth]{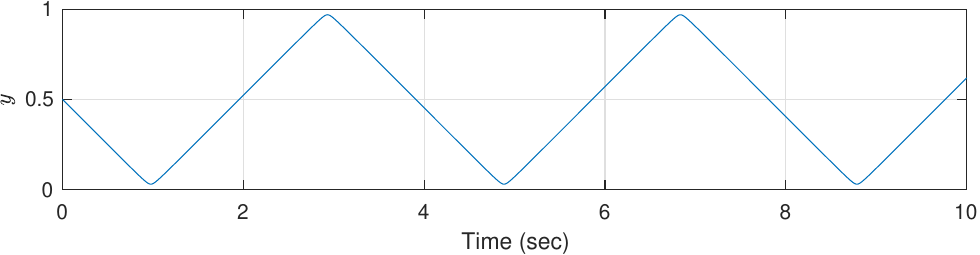}
    \caption{Trajectory with initial conditions $x(0)=0, y(0)=0.5, \dot x(0)=1, \dot y(0)=-0.5$; $\kappa=10^4$.}
\label{Figure22}
\end{figure}

\end{example}
\begin{example}{\bf Particle confined within a square.}
The same idea of the previous examples can be adapted also to cases where all the configuration coordinates are constrained. For instance let us consider the open square
$
M=\{(x,y)\in \R^2\; |\; 0<x<1, 0<y<1\}.
$
The Lagrangian with complete dynamics in $M$ becomes
\begin{eqnarray*}
L_{P''}&=&\left[\frac{1}{\kappa}\left(\frac{1}{y^2}+\frac{1}{ (y-1)^2}
+\frac{1}{ x^2}+\frac{1}{(x-1)^2}\right) -\frac{1}{2}\right] \left(\dot{x}^2+\dot{y}^2\right).
\end{eqnarray*}
\begin{figure}[h]
	\centering
 	\includegraphics[width=0.5\textwidth]{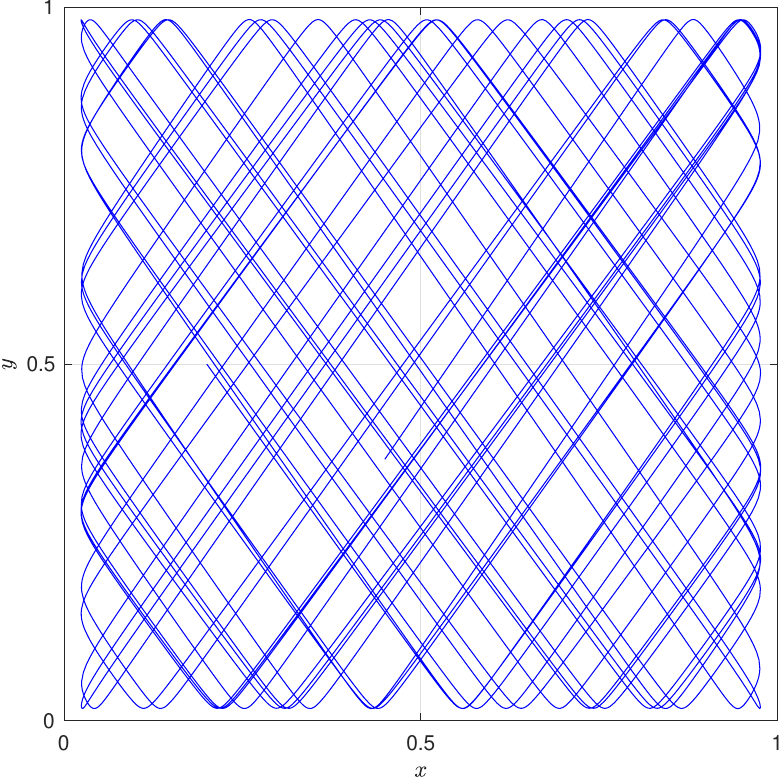}
\caption{Trajectory with initial conditions $x(0)=0.2$, $y(0)=0.5$, $\dot x(0)=0.5$, $\dot y(0)=-0.7$; $\kappa=10^4$.}
\label{Figure32}
\end{figure}
A path in the $x$-$y$ phase plane is shown in Fig. \ref{Figure32}.
\end{example}

\begin{example}{\bf Avoiding a circular obstacle.}
Consider the disk $D=\{(x, y)\in \R^2\; |\, x^2+y^2\leq 1\}$. We seek for a control strategy to avoid the disk obstacle finding a complete Riemannian  metric. For simplicity, assume that we have the following control system
\begin{equation}\label{disk-e}
\ddot{x}=x u, \; \qquad \ddot{y}= y u,
\end{equation}
and consider the complete Riemannian metric in $M={\mathbb R}^2\setminus D$ given by
\[
dx\otimes dx+dy\otimes dy + df\otimes df,
\]
where $f=\ln (x^2+y^2-1)$ (see Theorem \ref{teore}). 
Its corresponding Lagrangian reads
\[
L=\frac{1}{2} \left(\dot{x}^2+\dot{y}^2\right)+\frac{(8 x y) \dot{x} \dot{y}+\left(4 x^2\right) \dot{x}^2+4 y^2 \dot{y}^2}{2\left(x^2+y^2-1\right)^2},
\]
whose trajectories are complete in $M$. The Euler-Lagrange equations for $Lq$ gives us the control 
\[
u=\frac{2 \left(4 x y \dot{x} \dot{y}- y^2\dot{x}^2+x^2 \dot{x}^2+ \dot{x}^2-x^2 \dot{y}^2+ y^2 \dot{y}^2+ \dot{y}^2\right)}{\left(x^2+y^2-1\right) \left(30 x^2 y^2+15 x^4-28 x^2+15 y^4-28 y^2+15\right)}.
\]
A couple of symmetrical paths in the $x$-$y$ phase plane are shown in Fig. \ref{figurea1}.
\begin{figure}[H]
	\centering
    \includegraphics[scale=0.8]{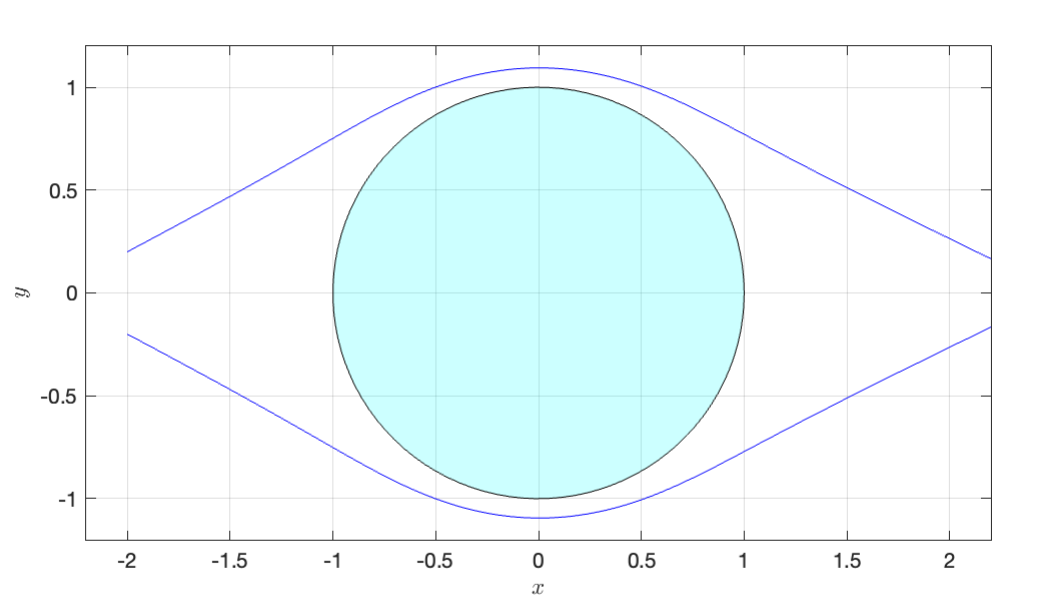}
\caption{Trajectory at the top of the circle with initial conditions $x(0)=-2, y(0)=0.2, \dot x(0)=1, \dot y(0)=0.53$; and its symmetrical with respect to axis $y=0$ at the bottom.}  
\label{figurea1}
\end{figure}
\end{example}
\begin{example}{\bf ``Bouncing control" on the interior of a disk.}
 Similarly to  Example  \ref{example-bou} take the Lagrangian function with complete trajectories in $D$
 \[
 L(x, y, \dot{x}, \dot{y})=\frac{1}{2} \left(1-\frac{1}{\kappa \left(x^2+y^2-1\right)^2}\right) \left(\dot{x}^2+\dot{y}^2\right),
 \]
whose  associated Euler-Lagrange equations are
\begin{eqnarray*}
 -\frac{4 y \dot{x}\dot{y}}{\kappa \left(x^2+y^2-1\right)^3}-\frac{2x  \left(\dot{x}^2-\dot{y}^2\right)}{\kappa \left(x^2+y^2-1\right)^3}+\ddot{x} \left(\frac{1}{\kappa \left(x^2+y^2-1\right)^2}-1\right)&=&0,\\
 -\frac{4 x \dot{x} \dot{y}}{\kappa \left(x^2+y^2-1\right)^3}+\frac{2 y \left(\dot{x}^2-\dot{y}^2\right)}{\kappa \left(x^2+y^2-1\right)^3}+\ddot{y} \left(\frac{1}{\kappa \left(x^2+y^2-1\right)^2}-1\right)&=&0.
\end{eqnarray*}
\begin{figure}
     \centering
     \includegraphics[width=\textwidth]{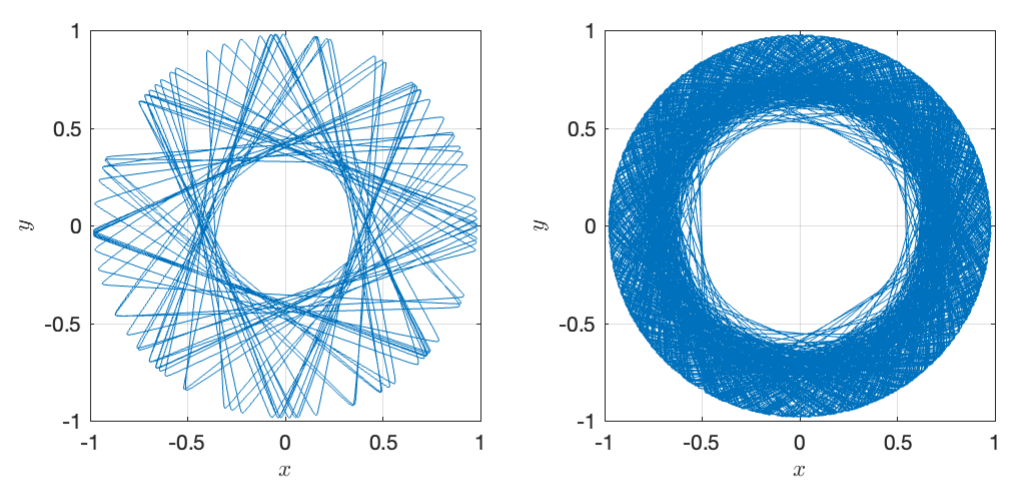}
    \caption{Trajectories with initial conditions:
left plot $x(0)=0.5, y(0)=0, \dot{x}(0)=0.1, \dot{y}(0)=0.1$; and 
right plot $x(0)=\frac{1}{4}$, $y(0)=\frac{\sqrt{3}}{4}$, $\dot{x}(0)=\frac{3}{2}$, $\dot{y}(0)=-\frac{\sqrt{3}}{2}$; $\kappa=10^3$. 
}
\label{figurea3-1}
\end{figure}
\end{example}
See Fig. \ref{figurea3-1} as an illustration of the trajectories which resemble in the limit an elastic collision with the boundary. 


\section{Constrained Controlled Lagrangians}\label{sec:constrained-control} 
In \cite{controlled-1} (see also \cite{controlled-2}) the authors developed a stabilization technique based on a modification of the Riemannian metric of the original mechanical Lagrangian where the extra-terms are identified as a control forces of the initial systems.
We will develop a general procedure to obtain a control strategy that additionally allows us to confine the motion of the system to a prescribed region determined by unilateral constraints in the configuration space.

Initially, we start with a system given by the following data $(g, V, {\mathcal F}_c)$ on a manifold $Q$ where $g$ is a Riemannian metric, $V$ a potential function and ${\mathcal F}_c$ the allowed control forces. 
We will assume that ${\mathcal F}_c=\hbox{span}\{ \mu_a, 1\leq a\leq m\}$ and by means of the Riemannian metric we define the control distribution ${\mathcal D}_c=\hbox{span}\{ Y_a=\sharp_g(\mu_a)\}$. The equations of motion for the mechanical control system are 
(\ref{control}):
$$
\nabla^g_{\dot{c}(t)}\dot{c}(t)+\hbox{grad}_g V(c(t))=u^a(t) Y_a(c(t)),
$$
which are locally expressed as in equations (\ref{eq:alterna-2a}) and (\ref{eq:alterna-2b}) and read
\begin{eqnarray}
(\mu_a)_k\left(\ddot{q}^k + {\Gamma}^k_{ij}\dot{q}^i\dot{q}^j+{g}^{ki}\frac{\partial V}{\partial q^i}\right)&=&{\mathcal C}_{ab}u^b, \label{eqr1}\\ 
  (\mu_\alpha)_k\left(\ddot{q}^k + {\Gamma}^k_{ij}\dot{q}^i\dot{q}^j+{g}^{ki}\frac{\partial V}{\partial q^i}\right)&=&0, \label{eqr2}
\end{eqnarray}
where $\mu_{\alpha}\in {\mathcal D}_c^0$, $n+1\leq \alpha\leq n$. That is $\mu_{\alpha}(Y_a)=0$ for all $1\leq a\leq m$.

Now, we use ideas similar to those in (\cite{controlled-1}) maintaining the  guiding principle of  considering a class of control laws that yield closed-loop dynamics which remain in Lagrangian form but fulfill the imposed unilateral constraint  applying Theorem \ref{teore}.
\begin{theorem}\label{teo-2}
Suppose we have  a Riemannian metric $\bar{g}$ satisfying the general matching conditions
\begin{equation}
   (\mu_\alpha)_k\left(\ddot{q}^k + \bar{\Gamma}^k_{ij}\dot{q}^i\dot{q}^j+\bar{g}^{ki}\frac{\partial V}{\partial q^i}\right)=0,  
\end{equation}
where $\bar{g}=\bar{g}_{ij}dq^i\otimes dq^j$ and $\bar{\Gamma}^k_{ij}$ are the corresponding Christoffel symbols. 
Let  $f: M\rightarrow {\mathbb R}$ be a proper function on an open subset  $M$ of $Q$ satisfying  $df\in \bar{\mathcal F}_c=\hbox{span}\{
\bar{\mu}_a=\flat_{\bar{g}}(Y_a)\}$ and define the control  
\begin{eqnarray*}
u^a_*&=&{\mathcal C}^{ab}(\mu_b)_k\left( \left[{\Gamma}^k_{ij}-\bar{\Gamma}^k_{ij}\right]\dot{q}^i\dot{q}^j+\left[{g}^{ki}-\bar{g}^{ki}\right]\frac{\partial V}{\partial q^i}\right)\\
&&+ \ \bar{\mathcal C}^{ab}(1+|\hbox{grad}_{\bar{g}} f|^2)^{-1}
\left(
\bar{f}^j\frac{\partial V}{\partial q^j}-\bar{f}_{jk}\dot{q}^j\dot{q}^k\right)\langle df, {Y}_b\rangle,
\end{eqnarray*}
where 
\begin{eqnarray*}
\bar{f}^i&=& \bar{g}^{ik}\frac{\partial f}{\partial q^k}, \\ 
|\hbox{grad}_{\bar{g}} f|^2
&=&\bar{g}^{ij}\frac{\partial f}{\partial q^i}
\frac{\partial f}{\partial q^j}, \\
\bar{f}_{jk}&=&\frac{\partial^2 f}{\partial q^j\partial q^k}-\bar{\Gamma}_{jk}^i\frac{\partial f}{\partial q^i}, \\
\bar{\mathcal C}_{ab}&=&\bar{g}(Y_a, Y_b),\quad (\bar{\mathcal C}_{ab})^{-1}=(\bar{\mathcal C}^{ab}).
\end{eqnarray*}
Then the system 
\begin{equation}\label{equ-controldetermined}
\nabla^g_{\dot{c}(t)}\dot{c}(t)+\hbox{grad}_g V(c(t))=u^a_*(c(t), \dot{c}(t)) Y_a(c(t))
\end{equation}
restricts the motion to $M$. 
\end{theorem}

\begin{proof}
The mechanical system determined by $(\bar{g}, V, {\mathcal D}_c)$ 
gives us the equations
\[
\ddot{q}^k+\bar{\Gamma}^k_{ij}\dot{q}^i\dot{q}^j+\bar{g}^{ki}\frac{\partial V}{\partial q^i}=\bar{u}^a {Y}_a^k, \quad 1\leq k\leq n,
\]
where ${Y}_a={Y}_a^i\frac{\partial}{\partial q^i}={g}^{ij}(\mu_a)_j\frac{\partial}{\partial q^i}$.
Given the proper function $f$ with $df\in \bar{\mathcal {F}}_c=\hbox{span}\{
\bar{\mu}_a=\flat_{\bar{g}}(Y_a)\}$ we can directly apply Theorem \ref{teore}
and we obtain the control 
\begin{equation}\label{equation:control-1}
\bar{u}^a=\bar{\mathcal C}^{ab}(1+|\hbox{grad}_{\tilde{g}} f|^2)^{-1}
\left(
\bar{f}^j\frac{\partial V}{\partial q^j}-\bar{f}_{jk}\dot{q}^j\dot{q}^k\right)\langle df, {Y}_b\rangle,
\end{equation}
where $\bar{\mathcal C}_{ab}=\bar{g}(Y_a, Y_b)$.
In consequence
 \begin{equation}\label{eqr3}
 (\mu_a)_k\left(\ddot{q}^k + \bar{\Gamma}^k_{ij}\dot{q}^i\dot{q}^j+\bar{g}^{ki}\frac{\partial V}{\partial q^i}\right)={\mathcal C}_{ab} \bar{u}^b\, . 
\end{equation}
Finally, subtracting equation (\ref{eqr1}) and (\ref{eqr3})  we obtain after some straightforward calculations the expression of the control $u_*^a$, and since the metric $\bar{g}+df\otimes df$ is complete the evolution is restricted to $M$. 
\end{proof}

In \cite{controlled-1} the authors modify a mechanical system of the type (\ref{control}) only using a modification of the initial Riemannian metric of the form
\[
\bar{g}(u_q, v_q)=g_{\sigma}(\hbox{Hor}_{\tau}u_q, \hbox{Hor}_{\tau}v_q)+g_{\rho}(\hbox{Ver}_{\tau}u_q, \hbox{Ver}_{\tau}v_q),
\]
where $u_q, v_q\in T_qQ$  where $\hbox{Hor}_{\tau}$ and $\hbox{Ver}_{\tau}$ are  particular choices of horizontal and vertical spaces and $g_{\sigma}$ and $g_{\rho}$ are metrics acting on these horizontal and vertical spaces, respectively. This produces the controlled Lagrangian determined by $(\bar{g}, V)$. 
The objective is using the Euler-Lagrange equations of the controlled Lagrangian to recover a control that stabilizes the initial system. 

Moreover, for asymptotic stability in \cite{controlled-1} it is necessary to add a dissipation to the system. The next proposition gives a way to introduce dissipation in Theorem \ref{teo-2}. 

\begin{proposition} \label{pr:diss}
The system given by  (\ref{equ-controldetermined}) becomes a dissipative system if we consider the following modified mechanical control system
\begin{equation}\label{equ-controldetermined-diss}
\nabla^g_{\dot{c}(t)}\dot{c}(t)+\hbox{grad}_g V(c(t))=\left[u^a_*+u^a_{dis}\right] Y_a(c(t))\; .
\end{equation}
where 
\begin{equation} \label{u-diss}
u^a_{dis}=(-(\bar{g}_f)_{ij}Y^i_a\dot{q}^j) 
\end{equation}
where $\bar{g}_f$ is the Riemannian metric $\bar{g}+df\otimes df$. Moreover, the new dissipative system remains complete. 
\end{proposition}
\begin{proof}
From Theorem \ref{teo-2} we know that the dynamics   (\ref{equ-controldetermined})   is equivalent to the free dynamics
given by $(\bar{g}_f=\bar{g}+df\otimes df, V)$, that is, 
   \begin{equation}\label{eq_lf+v}
\nabla^{\bar{g}_f}_{\dot{c}(t)}\dot{c}(t)+\hbox{grad}_{\bar{g}_f} V(c(t))=0\; .
\end{equation} 
Therefore for this dynamics the energy function
\[
E_{L_f}(v_q)=
\frac{1}{2}\bar{g}_f(v_q, v_q)+V(q)
\]
is a constant of the motion. Therefore adding the term
$
u^a_{dis}Y_a=(-(\bar{g}_f)_{ij}Y^i_a\dot{q}^j)  Y_a
$
we have  \begin{equation}\label{eq_lf+v}
\nabla^{\bar{g}_f}_{\dot{c}(t)}\dot{c}(t)+\hbox{grad}_{\bar{g}_f} V(c(t))=u^a_{dis}(c(t), \dot{c}(t))Y_a(c(t))\; .
\end{equation} 
and it is trivial to check that 
\begin{eqnarray*}
\frac{d E_{L_f}}{dt}&=&u^a_{dis}(c(t), \dot{c}(t))Y^i_a(c(t))(\bar{g}_f)_{ij})\dot{q}^j \nonumber \\
&=&
-\sum_{a=1}^m\left((\bar{g}_f)_{ij}Y^i_a\dot{q}^j\right)^2 
\leq 0.
\end{eqnarray*}
Therefore, the proof is a direct consequence of Theorem \ref{teo-2} and completeness from Theorem \ref{teo-principal}.
\end{proof}
The dissipation added can be written in a covariant way similar to equations (\ref{equ-controldetermined-diss}) as
\begin{equation} \label{u-diss-cov}
\frac{d}{dt}
\left(
\frac{\partial L}{\partial \dot{q}^i}\right)
-\frac{\partial L}{\partial {q}^i}=\left[u^a_*+u^a_{dis}\right] (\mu_a)_i,
\end{equation}
where $\mu_a=\flat_g(Y_a)$.

However, the energy function might not be locally convex around the equilibrium and therefore the dissipation as defined in the Proposition \ref{pr:diss} will not guarantee asymptotic stability as desired. Moreover, as it was shown in \cite{BLOCH199437} dissipation might induces instability, limit the region of stability \cite{WOOLSEY7076315} or even makes the equilibrium unstable as in the counterexample in the Hamiltonian setting in \cite{GOMEZESTERN2004955}. A way to somehow overcome this is to make use of dissipativity and define the target in terms of energy level set instead of the state equilibrium, say the desired energy level set $E_{L_f}^*$. In the following proposition we extend the previous dissipative result by constructing the appropriate composition.
\begin{proposition} \label{pr:diss_sg}
Consider the $C^{1}$ function $H \circ E_{L_f}(v_q)$ positive definite, proper and convex. The system given by (\ref{equ-controldetermined}) becomes a dissipative system if we consider the modified mechanical control system \eqref{equ-controldetermined-diss} with
\begin{equation} \label{u-diss-H}
u^a_{dis}=-\frac{\partial H}{\partial E_{L_f}} \cdot (\bar{g}_f)_{ij}Y^i_a\dot{q}^j,    
\end{equation}
where $\bar{g}_f$ is the Riemannian metric $\bar{g}+df\otimes df$. Moreover, the new dissipative system remains complete. 
\end{proposition}
\begin{proof}
The proof follows similar steps than the one of the Proposition \ref{pr:diss}. Thus, the derivative of the function $H$ along the solutions of the dynamics \eqref{equ-controldetermined-diss} becomes
\begin{eqnarray*}
\frac{d H}{dt}&=& \frac{d}{dt} H(E_{L_f}) \\
&=&\frac{\partial H}{\partial E_{L_f}} \frac{d E_{L_f}}{dt} \\
&=&\frac{\partial H}{\partial E_{L_f}} u^a_{dis}(c(t), \dot{c}(t))Y^i_a(c(t))(\bar{g}_f)_{ij})\dot{q}^j \\
&=& - \left( \frac{\partial H}{\partial E_{L_f}}\right)^2 \sum_{a=1}^m\left((\bar{g}_f)_{ij}Y^i_a\dot{q}^j\right)^2 \leq 0,
\end{eqnarray*}
where in the last line we have used \eqref{u-diss-H}. Therefore, the proof is a direct consequence of Theorem \ref{teo-2} and completeness from Theorem \ref{teo-principal}.
\end{proof}

In the next section, we show its use for stabilization of the pendulum on a cart at a prescribed energy level.

\subsection{Stabilization of a constrained pendulum on a cart}

Consider the underactuated pendulum on a cart (PoC) with Lagrangian function from \cite{Bloch} given by
\begin{align} \label{eq:Lo}
 L_o (\phi, \dot{\phi}, \dot{s})=\frac{1}{2} \left(\alpha \dot{\phi}^2 + 2\beta \cos(\phi) \dot{\phi} \dot{s} + \gamma\dot{s}^2 \right) + D \cos(\phi), 
\end{align}
with $\alpha$, $\beta$, $\gamma$, $D$ constant physical parameters and the subscript $(\cdot)_o$ stands for open loop.
Let us consider two different stabilization cases: a) the pendulum at the bottom position, and b) at the upright inverted position. The two  cases are substantially different because the equilibrium of the unforced case a) is stable ($D<0$), while the one of b) is unstable.

\subsubsection{Case a) stable equilibrium at $\phi=\pm\pi$.} 
Consider the control problem of maintaining the confinement of trajectories to the strip $S:=\{(\phi,s)\in S^1 \times \R| -1<s<1\}$ with a complete Riemannian metric. Thus, $M=\{s\in \R\; |\; -1<s<1\}$ and the proper map can be defined as $f(s):=\arcsin(s)$. 
The Lagrangian function with complete trajectories in $S$ reads
 \[
 \tilde L (\phi, s, \dot{\phi}, \dot{s}) := L_o (\phi, \dot{\phi}, \dot{s}) + \frac{1}{2} \frac{\dot{s}^2}{(1-s^2)}. 
 \]
 \begin{figure}[htbp]
	\centering
	\includegraphics[width=0.8\textwidth]{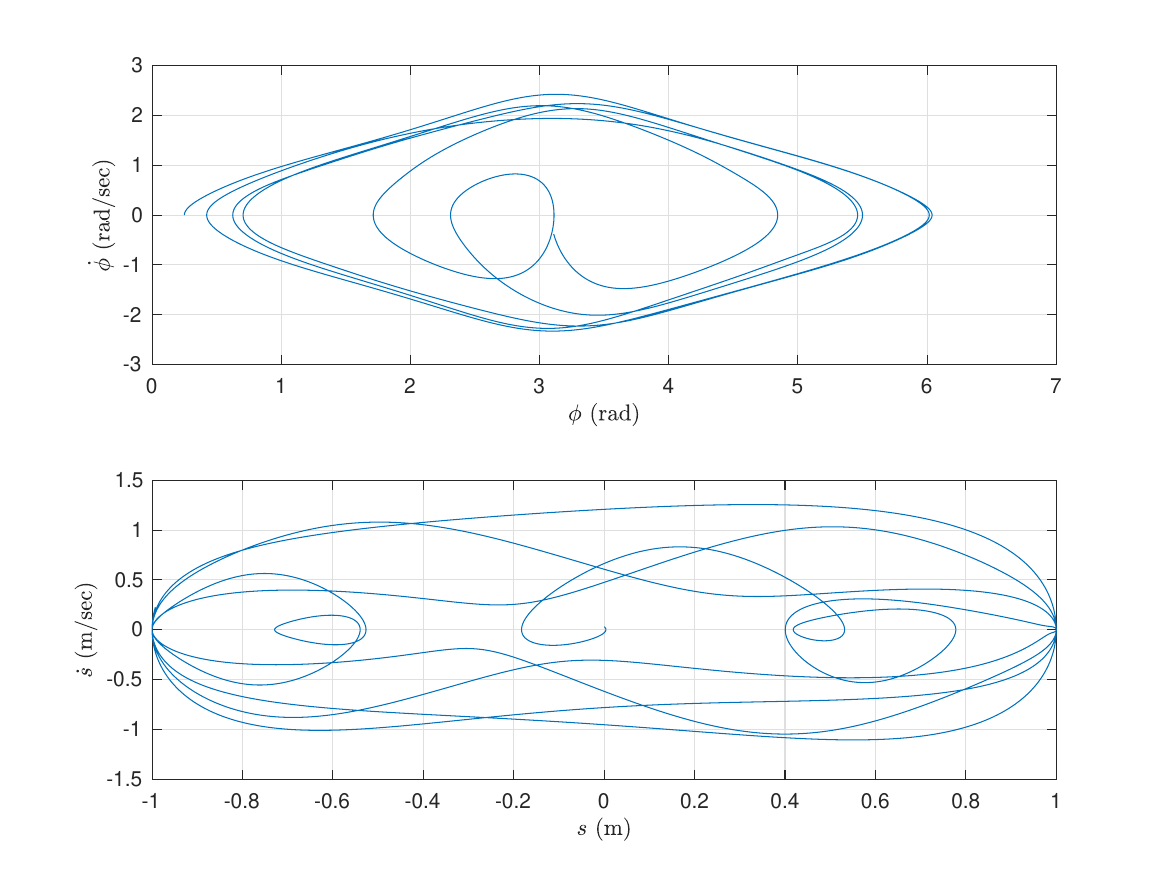}
    \caption{Bottom equilibrium with initial conditions $[\phi(0), s(0), \dot{\phi}(0), \dot{s}(0)]=[0.25, 0, 0, 0.03]$. See video file 
    \href{https://hdvirtual.us.es/discovirt/index.php/s/DfxKrGDPC7AZ7ed}{\underline{PoC down constrained}}.}
\label{poc_down}
\end{figure}

Define the key function $\rho:=\alpha/(|g_o| (1-s^2)) $ with $g_o$ is the metric defined in \eqref{eq:Lo} and $|g_o|=\textnormal{det}({g_o}_{ij})$  and the control reads
\begin{align}
    u = -\frac{\rho}{1 + \rho} \left[ \frac{\beta}{\alpha} \sin(\phi) \left(\alpha {\dot \phi}^2 + D \cos(\phi) \right)  + \frac{|g_o|}{\alpha} \frac{s}{1-s^2} {\dot s}^2 \right].
\end{align}

In Fig. \ref{poc_down} we show a typical simulation where notice how the controller enforces $|s|<1$.

\bigskip
\noindent{\bf Local Stability.}
Consider the state sorted as $x=\rm{col}(\phi, s, \dot \phi, \dot s)$, so that the first approximation at the equilibrium $x=x_\pi=\rm{col}(\pi, 0, 0, 0)$ can be described through the Jacobian matrix of the mechanical equations given by
\[
A_\pi = \frac{1}{|\tilde{g}_{x_\pi}|} \left[\begin{array}{cccc} 0 & 0 & 1 & 0 \\ 
0 & 0 & 0 & 1 \\
a_{31} & 0 & 0 & a_{34} \\
a_{41} & 0 & 0 & a_{44} \end{array}\right],
\]
where $|\tilde{g}_{x_\pi}|$ is the determinant of the metric tensor $\tilde{g}$ evaluated at $x_\pi$. 
Defining $\bar a_{ij}:=a_{ij}/|\tilde{g}_{x_\pi}|$, the characteristic polynomial in the variable $p$ reads
\[
 p \ (p^3 - \bar a_{44} p^2 - \bar a_{31} p + \bar a_{31} \bar a_{44} - \bar a_{34} \bar a_{41}) = 0.
\]
In the conservative case, i.e. without dissipation, the only nonzero coefficients are: $a_{31} = (1+\gamma) D$ and $a_{41} = \beta D$.
Hence, the roots are zero (double) and the imaginary $\pm\sqrt{a_{31}/|\tilde{g}_{x_\pi}|}$. The equilibrium is stable\footnote{A complete local stability analysis should include the Center Manifold provided by the zero eigenvalue.} and it corresponds to the simulation shown in the Fig. \ref{poc_down}.

\begin{figure}[htbp]
	\centering
 	\includegraphics[width=0.9\textwidth]{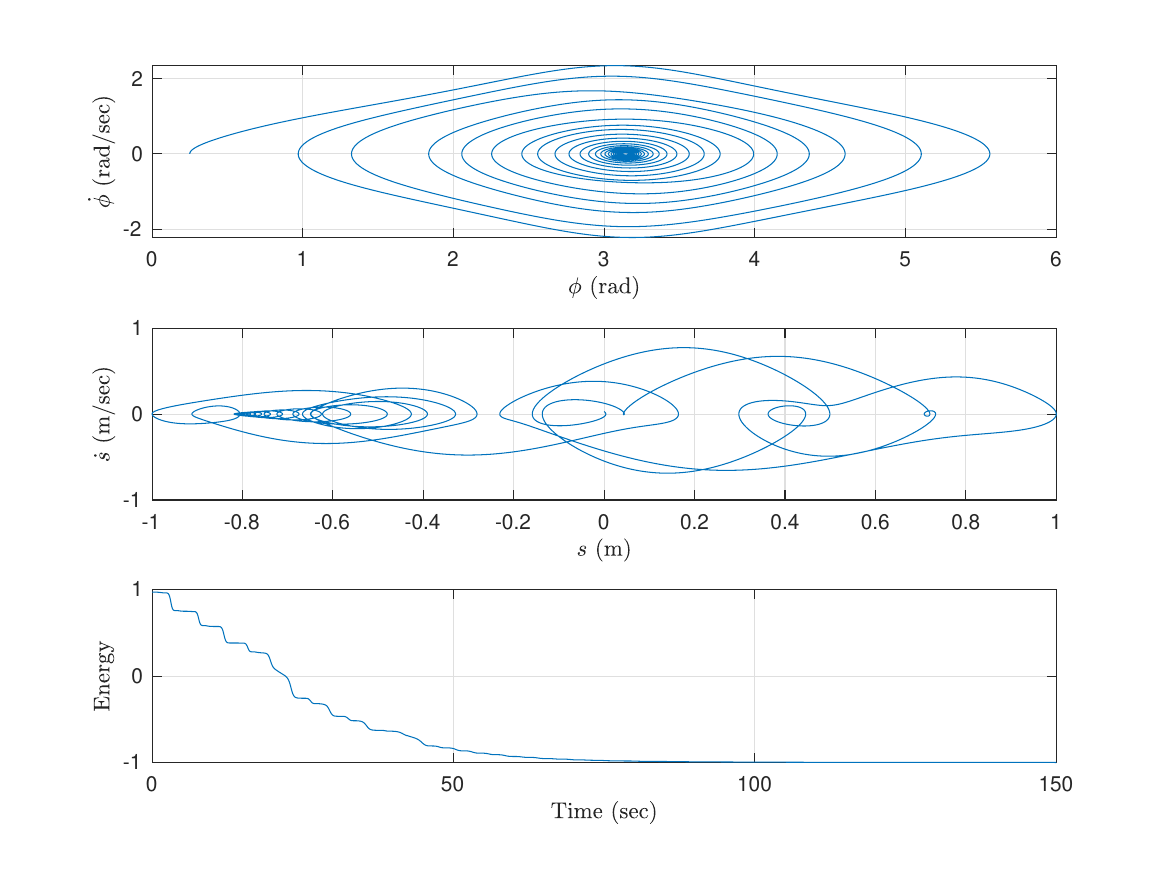}
\caption{Bottom equilibrium with initial conditions and dissipation $[\phi(0), s(0), \dot{\phi}(0), \dot{s}(0)]=[0.25, 0, 0, 0.03]$. Parameters: $k_{d}=0.3$. See video file 
\href{https://hdvirtual.us.es/discovirt/index.php/s/CXfYgZFW3Zpdw2M}{\underline{PoC down constrained and damped}}.}
\label{poc_down_damped}
\end{figure}

On the other hand, adding dissipation as defined in the Proposition \ref{pr:diss} given by \eqref{u-diss-cov} yields
$$u_{dis}=-k_d \dot s, $$ 
with $k_d>0$, and it enforces the dissipation inequality. Hence, the remaining terms of the Jacobian are: $a_{34} = - k_d \ \beta$, and $a_{44} = k_d \ \alpha$.
Thus, the Routh table for the characteristic polynomial yields stability for $k_d$ positive. A simulation is shown in the Fig. \ref{poc_down_damped}, where the energy time histories is shown at the bottom plot. This result is expected since the energy is locally convex around $x_\pi$ and hence it is a Lyapunov function.

\subsubsection{sCase b) stable equilibrium at $\phi=0$.} 
Now consider the control problem of maintaining confinement of the position of the  cart  while the pendulum is at a stable upright position. For this we need to modify the metric adding both a stabilizing term with $\bar{g}$ and the one for the constraint $d \bar f \otimes d \bar f $, i.e. shaping the total metric to $\tilde{g}=\bar{g} + d \bar f \otimes d \bar f$. Thus, for the stabilizing term we made use of the solution provided by controlled Lagrangians in \cite{controlled-1} where the Lagrangian is given by
 \[  
\bar L (\phi, \dot{\phi}, \dot{s})= L_o (\phi, \dot{\phi}, \dot{s}) + 
\frac{1}{2} \kappa (\kappa+1) \frac{\beta^{2}}{\gamma}\cos(\phi)^{2} \dot{\phi}^2 + \kappa \beta \cos(\phi) \dot{\phi} \dot{s},
\]
with $\kappa$ a constant parameter. In order to modify this metric $\bar g$ to add the constraint, recall that $d \bar f\in \bar{\mathcal F}_c=\hbox{span}\{\bar{\mu}_a=\flat_{\bar{g}}Y_a\}$, i.e. the forces induced by $\bar f$ must be in the $\hbox{span}\{\bar{\mu}_a\}$ which reads
\[
\bar{\mu}_a = \left[\begin{array}{c} \kappa \frac{\beta}{\gamma} \cos(\phi) \\ 1\end{array}\right].
\]
Thus, $\bar f_{k}=(\bar \mu_{a})_{k}$ can be integrated  and the general solution becomes $\bar f=\varphi(s + \kappa \frac{\beta}{\gamma} \sin(\phi))$, for some scalar function $\varphi(\cdot)$. This imposes a restriction on the class of constraints as it is shown below, but let us first introduce some notation for compactness as follows: $z:=s + \kappa \frac{\beta}{\gamma} \sin(\phi)$, $\bar \rho:=\frac{\delta}{|\bar g|} (\partial_{z} \varphi)^{2}$ and $\delta:= \alpha - \kappa \frac{\beta^{2}}{\gamma} \cos(\phi)^{2}$. Then, for any (smooth) $\varphi$ the general formula \eqref{equation:control-1} yields
%
%
\begin{equation*}
    \bar u = -\frac{\bar \rho}{1 + \bar \rho} \Bigg[\frac{\beta}{\delta} \sin(\phi) \left((\kappa + 1) \delta {\dot \phi}^2 + D \cos(\phi) \right) 
    - \kappa \frac{\beta}{\gamma} \sin(\phi) {\dot \phi}^2 \ (\partial_{z} \varphi)^{2} + \partial_{z} \varphi \ \partial_{z}^{2} \varphi \ (\dot s + \kappa \frac{\beta}{\gamma} \cos(\phi) {\dot \phi})^{2} \Bigg],
\end{equation*}
where $\partial_{z} \varphi := \frac{\partial \varphi(z)}{\partial z}$ and $\partial_{z}^{2} \varphi := \frac{\partial^{2} \varphi(z)}{\partial z^{2}}$.
The final Lagrangian can be written in coordinates as follows
\[  
\tilde L (\phi, s, \dot{\phi}, \dot{s}) =  \bar L (\phi, \dot{\phi}, \dot{s}) + \frac{1}{2} \left[\begin{array}{c}\dot{\phi} \\ \dot{s} \end{array}\right]^\top
\bar \mu_{a} \bar \mu_{a}^{\top} \left[\begin{array}{c}\dot{\phi} \\ \dot{s} \end{array}\right] (\partial_{z} \varphi)^{2}.
\]
The way to impose the constraint is indirectly through the still-free function $\varphi(z)$. Thus, let now the strip be defined as $M=\{z\in \R: |z|<r\}$ for some constant $r>0$. A proper map can be defined with $\varphi(z):=k_{b}\arcsin(z/r)$, $k_b>0$, which provides a bound for the position $s$ because $|z|=|s+\kappa \beta \sin(\varphi)/\gamma|<r$ implies $\max\{|s|\} \leq r + \kappa \beta/\gamma$, with $r$ free.

\begin{figure}[htbp]
	\centering
    \includegraphics[width=0.49\textwidth]{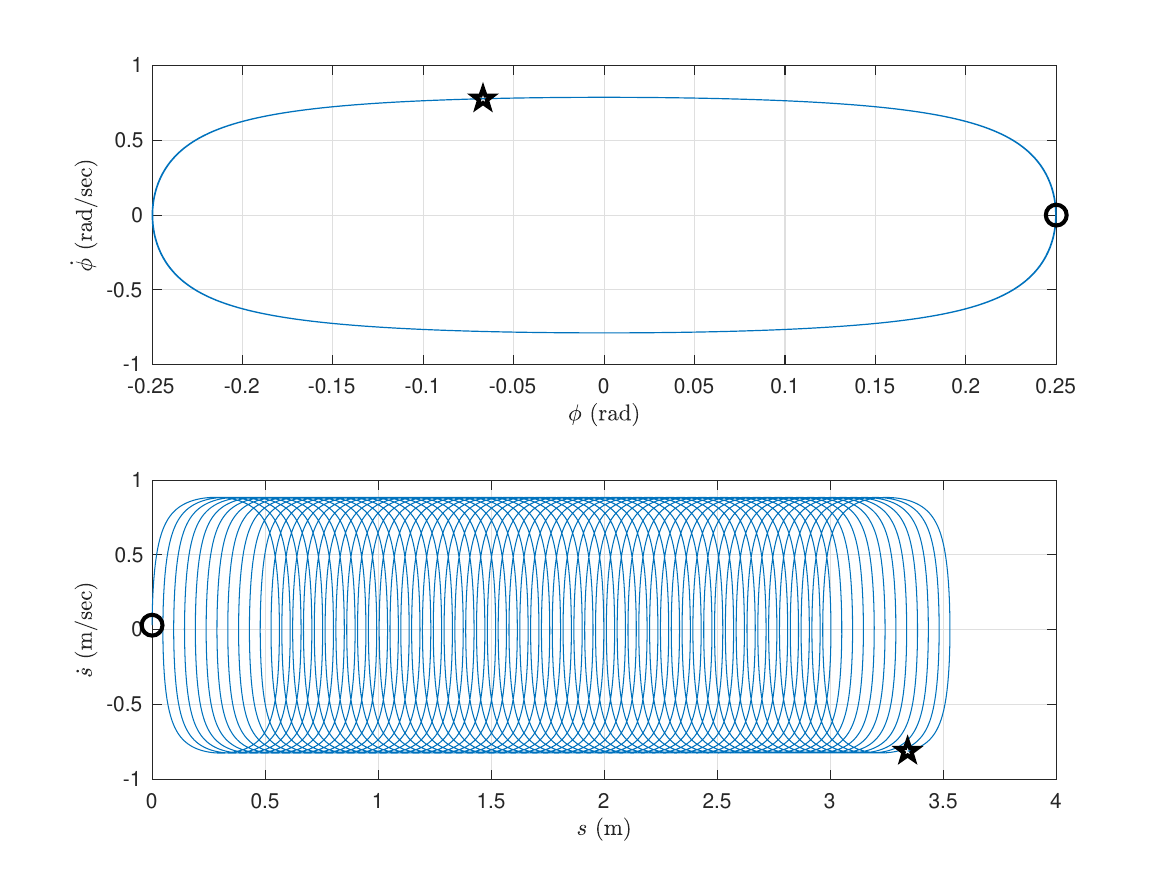} 
    \includegraphics[width=0.49\textwidth]{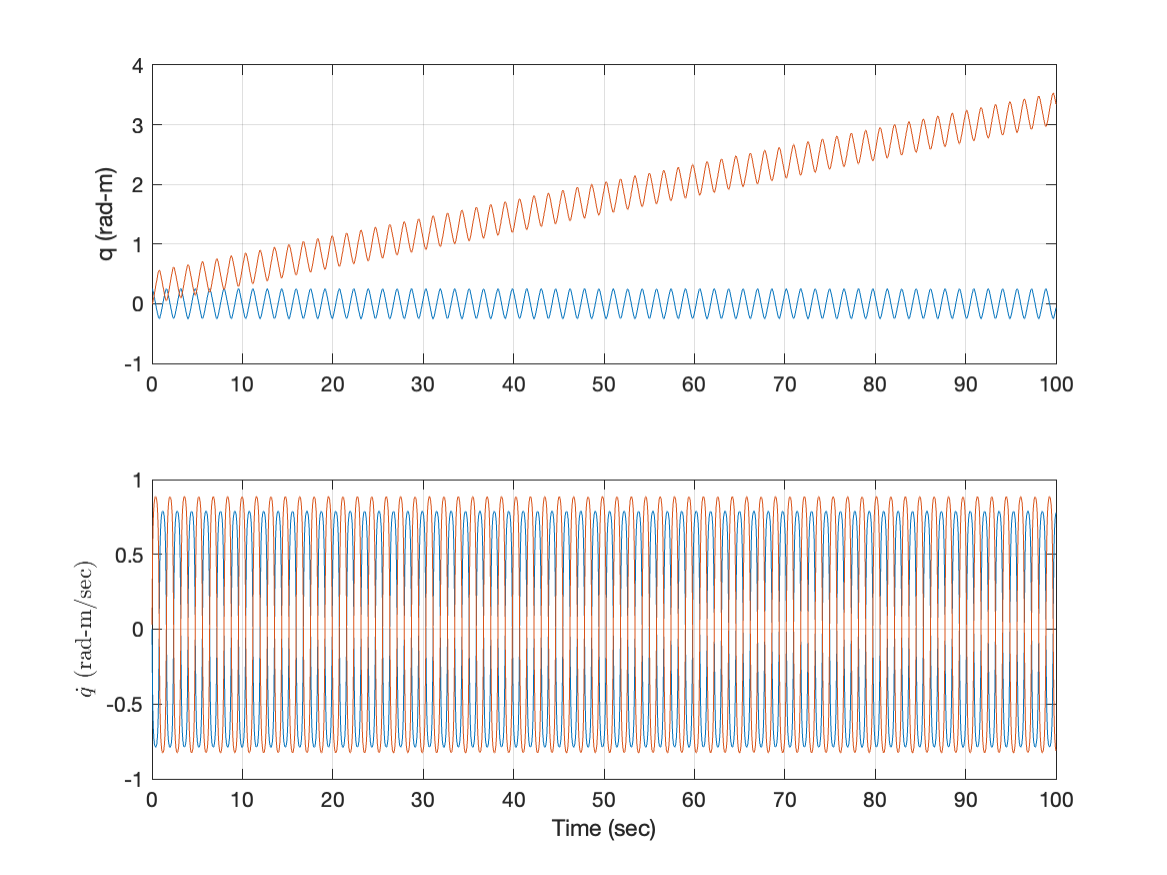}
\caption{Upright inverted position with initial conditions $[\phi(0), s(0), \dot{\phi}(0), \dot{s}(0)]=[0.25, 0, 0, 0.03]$ without constrained control. Parameters: $k_{b}=0$ and $\kappa=1.2$. See video file 
\href{https://hdvirtual.us.es/discovirt/index.php/s/kzacEk4t9R3CfLi}{\underline{PoC Unconstrained}}.}
\label{poc_up_u}
\end{figure}

\begin{figure}[htbp]
	\centering
    \includegraphics[width=0.49\textwidth]{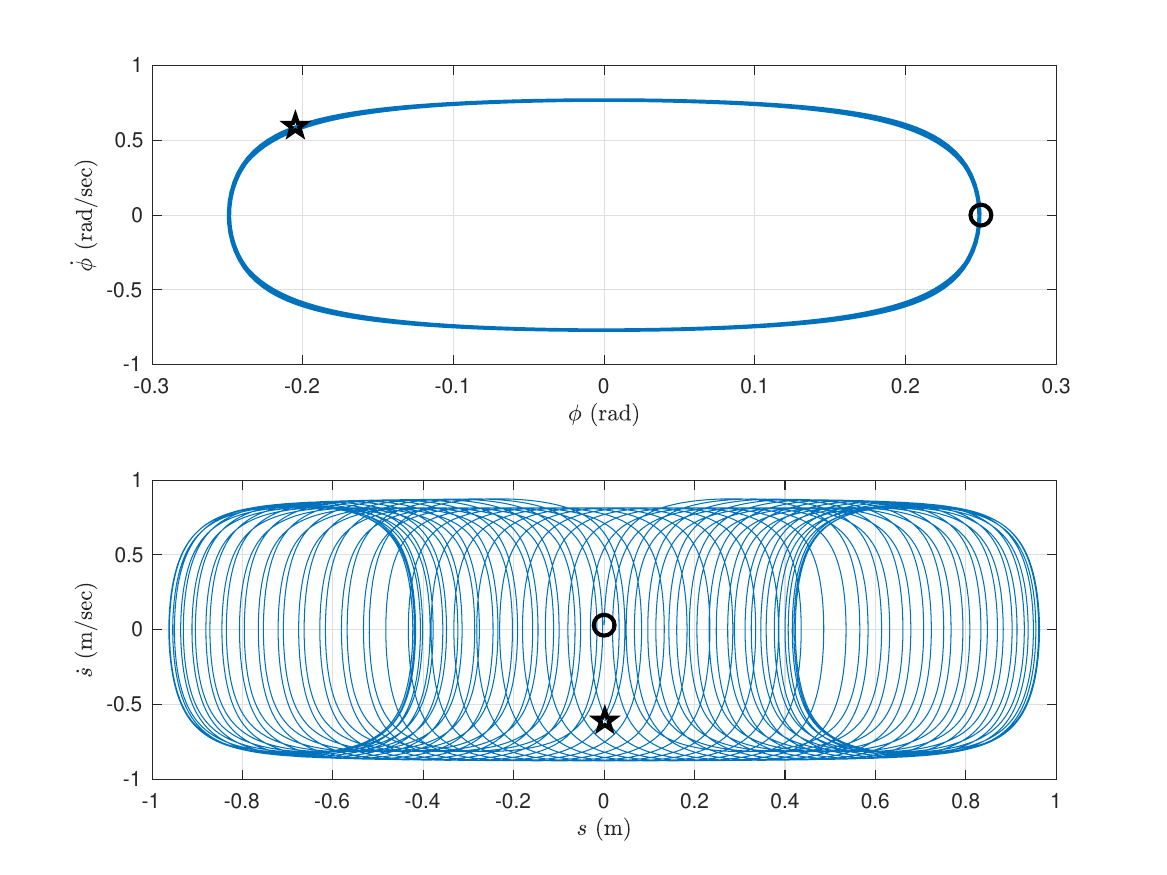} 
    \includegraphics[width=0.49\textwidth]{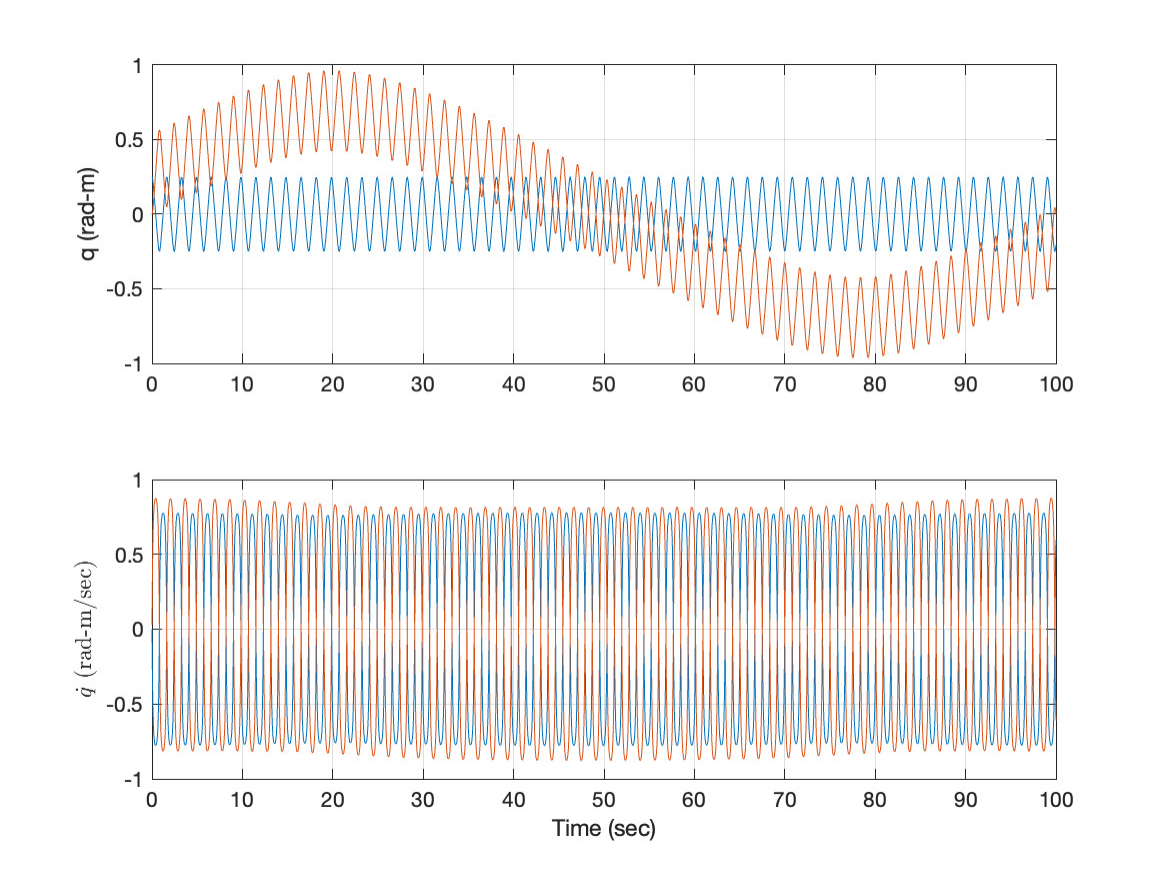}
\caption{Upright inverted position with initial conditions $[\phi(0), s(0), \dot{\phi}(0), \dot{s}(0)]=[0.25, 0, 0, 0.03]$ with constrained control. Parameters: $\kappa=1.2$, $k_{b}=0.1$ and $r=1$. See video file 
\href{https://hdvirtual.us.es/discovirt/index.php/s/3eWLgLxYnAf5mBJ}{\underline{PoC Constrained}}.}
\label{poc_up_c}
\end{figure}

To better explain this case we provide a couple of simulations in Figures \ref{poc_up_u} and \ref{poc_up_c}. Both simulations have the same initial conditions but the simulation of Fig. \ref{poc_up_u} is made only with the Controller Lagrangians, i.e. without the constrained controller with $k_{b}=0$, and that of Fig. \ref{poc_up_c} adds the constrained controller, where the bound for $s$ is satisfied such that the CoG of the PoC does not cross the boundary.

\bigskip
\noindent{\bf Local Stability.}
Consider again the state sorted as $x=\rm{col}(\phi, s, \dot \phi, \dot s)$, so that the first approximation at the equilibrium $x=x_0=\mathbf{0}$ can be described through the Jacobian matrix of the mechanical equations given by
\[
A_o = \frac{1}{|\tilde{g}_{x_0}|} \left[\begin{array}{cccc} 0 & 0 & 1 & 0 \\ 
0 & 0 & 0 & 1 \\
a_{31} & 0 & a_{33} & a_{34} \\
a_{41} & 0 & a_{43} & a_{44} \end{array}\right],
\]
where $|\tilde{g}_{x_0}|$ is the determinant of the metric tensor $\tilde{g}$ evaluated at $x_0$. 
\[
|\tilde{g}_{x_0}| = |\bar{g}_{x_0}| + \delta(0) {k_b^4}/{r^4}
\]

Defining $\bar a_{ij}:=a_{ij}/|\tilde{g}_{x_0}|$, the characteristic polynomial in the variable $p$ reads
\[
 p \ (p^3 - (\bar a_{33} + \bar a_{44}) p^2 + (\bar a_{33} \bar a_{44} - \bar a_{34} \bar a_{43} - \bar a_{31}) p + \bar a_{31} \bar a_{44} - \bar a_{34} \bar a_{41}) = 0.
\]
In the conservative case, i.e. without dissipation, the only nonzero coefficients are also $a_{31}$ and $a_{41}$ which are
\begin{equation*}
    a_{31} = -\gamma D \left(1+\frac{k_b^4}{\gamma r^4}\right), \quad
    a_{41} = \beta D \left[1+ \kappa \left(1 + \frac{k_b^4}{\gamma r^4}\right) \right].
\end{equation*}
Hence, the roots are a zero (double) and the imaginary $\pm\sqrt{a_{31}/|\tilde{g}_{x_0}|}$. Note that equilibrium is stable and it corresponds to the simulation shown in the Fig. \ref{poc_up_c}. Also notice that for $k_b=0$, i.e. no constraint, we recover the stability of the controlled Lagrangians.

On the other hand, adding dissipation as defined in the Proposition \ref{pr:diss} given by \eqref{u-diss} yields
$$u_{dis}=-k_d  \left(\dot s + \kappa \frac{\beta}{\gamma} \cos(\phi) \dot \phi \right), $$ 
with $k_d$ a constant, enforcing then dissipativity. Hence, the remaining terms of the Jacobian are: $a_{33} = -k_d \kappa {\beta^2}/{\gamma}$, $a_{34} = - k_d \beta$, $a_{43} = k_d  \kappa {\alpha \beta}/{\gamma}$ and $a_{44} = k_d  \alpha$.
%
%
The Routh table for the characteristic polynomial provides instability for any $k_d$, since there is an unstable manifold passing through the critical point. This result is expected since the energy is locally non-convex around $x_o$, and hence by extracting or injecting energy with \eqref{u-diss} the trajectories do not rest anywhere. 


However, as it was aforementioned a way to somehow overcome the lack of convexity of energy functions is the use of the dissipation inequality provided by a positive definite storage function as defined in Proposition~\ref{pr:diss_sg}.
\begin{figure}[htbp]
	\centering
	\includegraphics[width=0.45\textwidth]{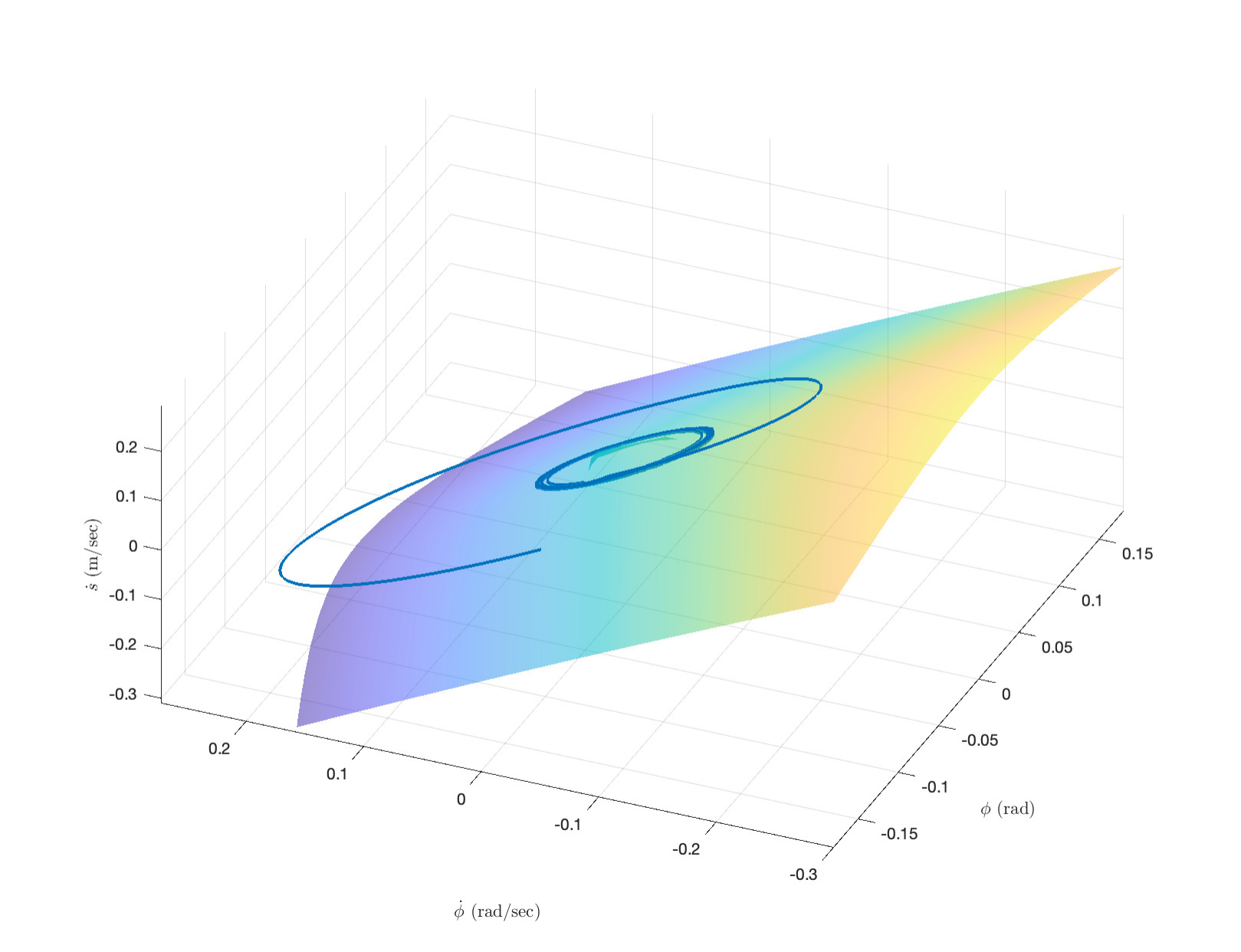} \hfill
    \includegraphics[width=0.5\textwidth]{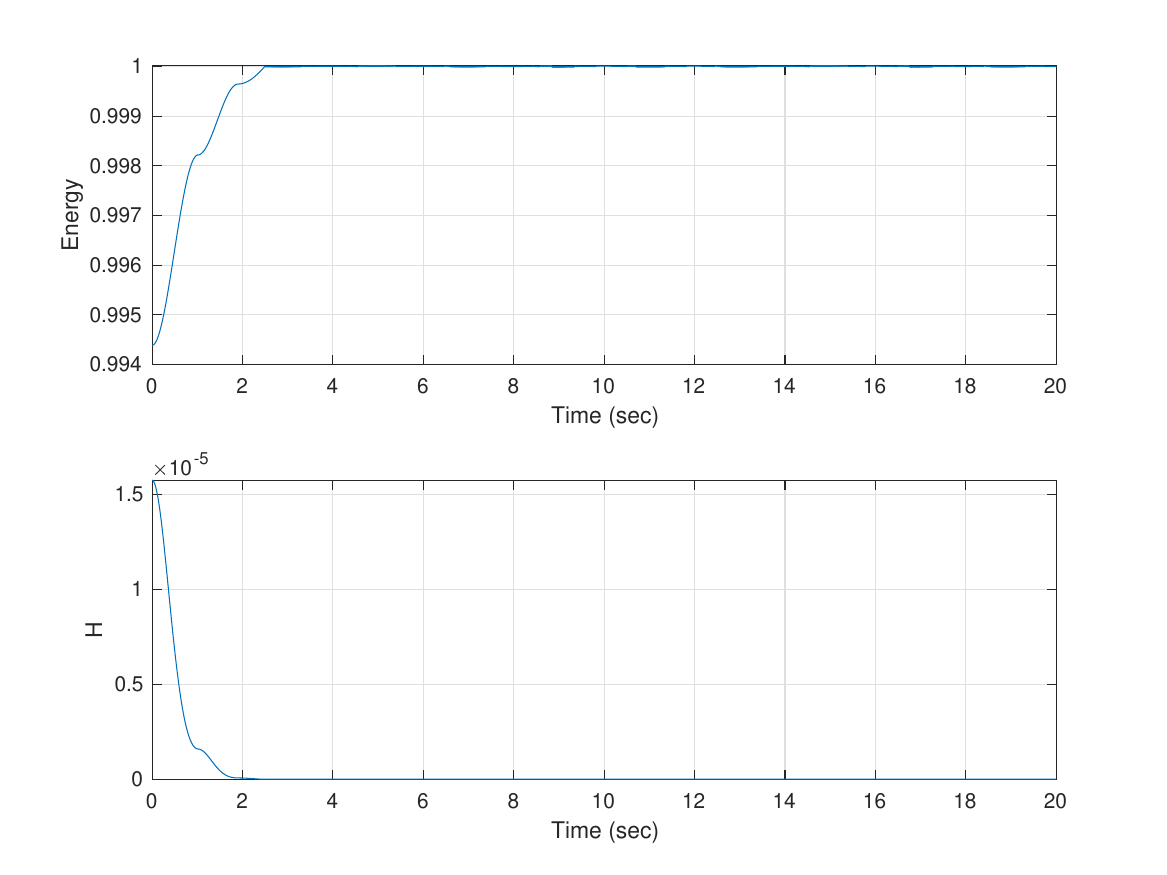}
\caption{The surface is the energy level set $E_{L_f}^*=1$ for the pendulum on a cart example. The trajectory converges to $E_{L_f}^*$ that corresponds to stable oscillations of the inverted pendulum.}
\label{poc_sg}
\end{figure}
Thus, let us add dissipation with the storage function $H=(E_{L_f}-E_{L_f}^*)^2/2$ with $E_{L_f}^*$ the desired energy level set. This storage function defines the dissipative control part as \eqref{u-diss-H} that becomes

$$u_{dis}=-k_d  (E_{L_f}-E_{L_f}^*) \left(\dot s + \kappa \frac{\beta}{\gamma} \cos(\phi) \dot \phi \right). $$

In Fig. \ref{poc_sg} left we show the trajectory of the pendulum on a cart from its initial condition $E_{L_f}\neq 1$. The time histories of both functions $E_{L_f}$ and $H$ are shown in Fig. \ref{poc_sg} right, where note that the energy  $E_{L_f}$ approaches rapidly to the desired level set $E_{L_f}^*=1$. Notice that the level set $E_{L_f}=1$ includes the inverted position.



\section{Conclusions and future work}\label{section:conclusions}
We have shown how the theory of incomplete-complete  metrics can be used in the control of constrained mechanical systems.
In addition we have shown how to combine the theory with the theory of controlled
Lagrangians to introduce the study of  stabilization and obstacle avoidance for the control of constrained-mechanical problems. 
In future work, we intend to extend these ideas to 
time-dependent systems,   systems with nonholonomic constraints, optimal control problems in 
the presence of obstacles and interpolation problems. 

\section*{Acknowledgements}
JÁ.A. was partially supported by the Project HOMPOT, grant number P20\_00597, under framework PAIDI 2020 European Commission FEDER funding. DMdD  acknowledges financial support from the Spanish Ministry of Science and Innovation, under grants  PID2022-137909NB-C21, TED2021-129455B-I00 and  CEX2019-000904-S funded by MCIN/AEI\-/10.13039\-/501100011033 and   BBVA Foundation via the project “Mathematical optimization for a more efficient, safer and decarbonized maritime transport”.
 A.B. was partially supported by NSF grant  DMS-2103026, and AFOSR grants FA
9550-22-1-0215 and FA 9550-23-1-0400.

\bibliography{References}


\end{document}